\documentclass{elsart}
\usepackage{graphics}
\usepackage{amssymb}
\journal{Physics Letters B}

\begin{document}
\begin{frontmatter}

\title{Measurement of the quadratic slope parameter in the $ K_{L}\rightarrow 3\pi ^{0} $ decay Dalitz plot}

%\collab{NA48 Collaboration}

\author{A.~Lai},
\author{D.~Marras},
\author{L.~Musa}
\address{Dipartimento di Fisica dell'Universit\`{a} e Sezione dell'INFN
di Cagliari, I-09100 Cagliari, Italy}

%\author{R.~Batley},
\author{A.~Bevan},
\author{R.S.~Dosanjh},
\author{T.J.~Gershon\thanksref{kek}},
\author{B.~Hay\thanksref{cern}},
\author{G.E.~Kalmus\thanksref{kalmus}},
\author{D.J.~Munday},
\author{M.D.~Needham\thanksref{needham}},
\author{E.~Olaiya},
\author{M.A.~Parker},
\author{T.O.~White},
\author{S.A.~Wotton}
\thanks[kek]{Present address: High Energy Accelerator Research
Organization (KEK), Tsukuba, Ibaraki, 305-0801, Japan}
\thanks[cern]{Present address: EP, CERN, CH-1211, Geneva 23, Switzerland}
\thanks[kalmus]{Based at Rutherford Appleton Laboratory, Chilton, Didcot,OX11 0QX, U.K}
\thanks[needham]{Present address: NIKHEF, PO Box 41882, 1009 DB Amsterdam, The Netherlands}
\address{Cavendish Laboratory, University of Cambridge, Cambridge, CB3
0HE, U.K\thanksref{cambridge}}
\thanks[cambridge]{Funded by the U.K. Particle Physics and Astronomy Research Council}

% ...

\begin{abstract}
A value of \( (-6.1\pm 0.9_{stat}\pm 0.5_{syst})\times 10^{-3} \) is obtained
for the quadratic slope parameter \( h \) in the \( K_{L}\rightarrow 3\pi ^{0} \)
decay Dalitz plot at the NA48 experiment at the CERN SPS. The result is based
on \( 14.7\times 10^{6} \) fully reconstructed \( K_{L}\rightarrow 3\pi ^{0}\rightarrow 6\gamma  \)
decays. This is the most precise measurement of any of the Dalitz plot slope
parameters in the charged and neutral kaon decays so far. 
\end{abstract}

\end{frontmatter}

\section*{Introduction}

The \( K\rightarrow 3\pi  \) decays Dalitz plot distributions can be expanded
in powers of Dalitz plot variables \( u \) and \( v \) \cite{PDG}:

{\par\raggedright \begin{equation}
\label{dalitz_plot}
\left| M\left( u,v\right) \right| ^{2}\propto 1+gu+jv+hu^{2}+kv^{2}
\end{equation}
 \par}

\[
u=\frac{\left( s_{3}-s_{0}\right) }{m^{2}_{\pi ^{+}}},\: v=\frac{\left( s_{1}-s_{2}\right) }{m^{2}_{\pi ^{+}}}\]
\[
s_{0}=\frac{s_{1}+s_{2}+s_{3}}{3},\: \: s_{i}=\left( P_{K}-P_{i}\right) ^{2},\: i=1,2,3\]
where \( P_{K} \) and \( P_{i} \) are the four-momenta of the decaying kaon
and \( i- \)th pion (\( i=3 \) for the {}``odd charge{}'' pion), respectively.

In the case of the \( K_{L}\rightarrow 3\pi ^{0} \) decay the expression (\ref{dalitz_plot})
reduces to \cite{Devlin}:

\begin{equation}
\label{dalitz_plot_k3pi0}
\left| M_{000}\left( R^{2},\theta \right) \right| ^{2}\propto 1+hR^{2}\: ,
\end{equation}

\[
R^{2}=\left( u^{2}+v^{2}/3\right) ,\: \theta =\arctan \left( v/\sqrt{3}u\right) \]
because of the identical final state particles. A positive/negative value of
the quadratic slope parameter \( h \) would mean that asymmetric/symmetric
final states are favoured.

Based only on isospin symmetry and some general assumptions it is possible to
define a relation \begin{equation}
\rho \equiv h+3k-\frac{g^{2}}{4\cos ^{2}\beta }
\end{equation}
 between the quadratic \( h,k \) and linear \( g \) Dalitz plot slope parameters
and the final state interaction phases \( \beta  \) which should be approximately
the same for all \( K\rightarrow 3\pi  \) decays \cite{Belkov1}. For \( K_{L}\rightarrow 3\pi ^{0} \)
decays the parameter \( \rho  \) is equal to \( 2\times h \) (\( k\equiv h/3 \))
and is not influenced by radiative corrections since all the final state particles
are neutral. Thus a high statistics measurement of the quadratic slope parameter
in the \( K_{L}\rightarrow 3\pi ^{0} \) decay Dalitz plot could establish strong
constraints to the Dalitz plot slope parameters and f.s.i. phases in other \( K\rightarrow 3\pi  \)
decays. Also, it could be useful in the evaluation of radiative corrections
applied to \( K^{\pm }\rightarrow \pi ^{\pm }\pi ^{\pm }\pi ^{\mp } \) and
\( K_{L}\rightarrow \pi ^{+}\pi ^{-}\pi ^{0} \) decays. On the other hand,
by combining the \( K_{L}\rightarrow 3\pi ^{0} \) decay Dalitz plot slope parameter
\( h \) with the linear and quadratic slope parameters in the \( K^{\pm }\rightarrow \pi ^{\pm }\pi ^{\pm }\pi ^{\mp } \)
decay it is possible to probe the validity of the \( \Delta I=1/2 \) rule.

The \( K_{L}\rightarrow 3\pi ^{0} \) Dalitz plot slope parameter \( h \) has
been estimated in the framework of ChPT \cite{Kambor1},\cite{Kambor2},\cite{Ambrosio}.
The analysis includes a nonzero \( \Delta I=3/2 \) amplitude in the quadratic
term of the Dalitz plot expansion (\ref{dalitz_plot}) by taking into account
the next-to-leading \( O(p^{4}) \) corrections. The phenomenological coupling
constants in the chiral Lagrangian are evaluated using a fit to the recent experimental
data in both \( K\rightarrow 2\pi  \) and \( K\rightarrow 3\pi  \) decays.

In this paper we present a measurement based upon a data set collected with
the NA48 detector in 1998. The detector acceptance has been estimated using
a detailed Monte Carlo (MC) simulation.

\section*{Detector setup and data taking}

The NA48 experiment is dedicated to the measurement of direct CP violation in
\( K_{L,S}^{0}\rightarrow 2\pi  \) decays. The \( K_{L} \) beam is produced
by \( 450 \) GeV/c protons from the CERN SPS hitting a \( 2 \) mm diameter
and \( 400 \) mm long beryllium target at a production angle of \( 2.4 \)
mrad. Passing through a set of collimators, a \( \pm 0.15 \) mrad divergent
\( K_{L} \) beam enters the decay volume \( 126 \) m downstream of the target.
The evacuated \( 89 \) m long decay volume is followed by a helium tank which
contains four drift chambers of the charged particle spectrometer. After the
tank a scintillator hodoscope, a liquid krypton electromagnetic calorimeter,
a hadron calorimeter and muon veto counters are placed.

The liquid krypton calorimeter (LKr) is designed to measure the energy, position
and timing of electromagnetic showers \cite{LKr}. The \( 127 \) cm long detector
consists of 13212 readout cells in a projective tower geometry which points
to the middle of the decay volume. Each cell of \( 2\times 2 \) cm\( ^{2} \)
cross section is made of copper-beryllium ribbons which are extended longitudinally
in a \( \pm 48 \) mrad accordion structure. The readout cells are contained
in a cryostat filled with about \( 20 \) t liquid krypton at \( 121 \) K.
The initial current induced on the copper-beryllium electrodes is measured by
\( 80 \) ns FWHM and digitized by \( 40 \) MHz FADCs. For the 1998 data taking
period, the energy and position resolutions of the calorimeter were determined
to be:

\begin{equation}
\label{e_{r}esol}
\frac{\sigma (E)}{E}=\frac{(3.2\pm 0.2)\%}{\sqrt{E}}\oplus \frac{(0.09\pm 0.01)}{E}\oplus (0.42\pm 0.05)\%
\end{equation}

\begin{equation}
\label{x_{r}esol}
\sigma (x)\simeq \sigma (y)\simeq \frac{0.4\: \rm {cm}}{\sqrt{E}}\oplus 0.05\: \rm {cm}
\end{equation}

with \( E \) measured in GeV.

The time resolution was better than \( 300 \) ps for photons with energies
above \( 20 \) GeV.

The energy non-linearity in the calorimeter response was found to be less than
\( 1\times 10^{-3} \) between \( 6 \) GeV and \( 100 \) GeV \cite{Unal}.

The LKr detector contains the so-called neutral hodoscope which is made of a
4 mm thick plane of scintillating fibres placed near the maximum of the e.m.
shower development. The neutral hodoscope is used in the estimation of the main
trigger inefficiency and for an additional measurement of the event time.

A description of the whole experimental setup can be found elsewhere \cite{setup}.

The \( K_{L}\rightarrow 3\pi ^{0}\rightarrow 6\gamma  \) data sample was acquired
by three different triggers. The first one is the neutral \( 2\pi ^{0} \) trigger
(NUT \( 2\pi ^{0} \)) in which each \( 25 \) ns the calibrated analogue sums
of \( 2\times 8 \) LKr cells are used to construct \( 64 \)-channel \( x \)
and \( y \) projections of the energy deposited in the calorimeter \cite{NUT}.
Based on these projections the total deposited energy \( E_{sum} \), the first
and second moments of the \( x \) and \( y \) energy distributions, and the
time of the energy peaks are computed. The centre-of-gravity COG of the event
and the longitudinal vertex position are reconstructed as: \begin{equation}
\label{cog_{t}rig}
COG^{trig}=\frac{\sqrt{M^{2}_{1,x}+M^{2}_{1,y}}}{E_{sum}}
\end{equation}
\begin{equation}
\label{z_{t}rig}
z^{trig}_{vertex}=z_{LKr}-\frac{\sqrt{E_{sum}\left( M_{2,x}+M_{2,y}\right) -\left( M^{2}_{1,x}+M^{2}_{1,y}\right) }}{m_{K}}
\end{equation}
 where \( M_{1,x} \) , \( M_{1,y} \) and \( M_{2,x} \), \( M_{2,y} \) are
the first and second moments of the energy deposition, \( z_{LKr} \) is the
longitudinal position of the calorimeter and \( m_{K} \) is the nominal kaon
mass.

The proper decay time \( \tau ^{trig} \) from the beginning of the decay region
(just after the final collimator) in units of the \( K_{S} \) lifetime \( \tau _{S} \)
is derived taking into account calibration constants. The events are accepted
by the NUT \( 2\pi ^{0} \) trigger if \( E_{sum}>50 \) GeV, \( COG^{trig}<15 \)
cm and \( \tau ^{trig}/\tau _{S}<5 \). In addition, less than \( 6 \) energy
peaks within \( 13.5 \) ns in both the \( x \) and \( y \) projections are
required.

The second trigger (NUT \( 3\pi ^{0} \)) uses the same hardware chain as NUT
\( 2\pi ^{0} \) trigger but is specially set-up to select \( 3\pi ^{0} \)
events by applying no condition on the number of the energy peaks.

The third trigger (Nhodo) is based on the information from the neutral hodoscope.
It requires a coincidence of signals from the upper and lower or the left and
right parts of the calorimeter. In order to reduce the trigger output rates
the Nhodo trigger and the NUT \( 3\pi ^{0} \) one were properly down-scaled.

\section*{Data analysis}

The \( K_{L}\rightarrow 3\pi ^{0}\rightarrow 6\gamma  \) events were selected
from all events which met at least one of the trigger requirements. The selection
criteria required 6 or more LKr clusters satisfying the following conditions:

\begin{itemize}
\item The energy of the cluster had to be between \( 3 \) GeV and \( 100 \) GeV; 
\item To avoid energy losses in the LKr, the distance from the cluster to the closest
dead calorimeter cell was required to be greater than \( 2 \) cm and the cluster
had to be more than \( 5 \) cm from the beam pipe and the outer edge of the
calorimeter. 
\end{itemize}
All possible combinations of \( 6 \) clusters which passed these requirements
were considered. In addition, the following further selection criteria were
applied on each combination:

\begin{itemize}
\item The distance between all cluster pairs had to be greater than \( 10 \) cm in
order to avoid overlapping of the clusters; 
\item All \( 6 \) clusters had to lie within \( 5 \) ns from the average cluster
time; 
\item The sum of the clusters energies had to exceed \( 60 \) GeV which is sufficiently
far from the NUT \( 2\pi ^{0} \) and NUT \( 3\pi ^{0} \) trigger threshold
of \( 50 \) GeV; 
\item The radial position of the center-of-gravity\\
\[
COG=\frac{\sqrt{\left( \sum ^{6}_{i=1}E_{i}x_{i}\right) ^{2}+\left( \sum ^{6}_{i=1}E_{i}y_{i}\right) ^{2}}}{\sum ^{6}_{i=1}E_{i}}\: ,\]
where \( E_{i} \), \( x_{i} \), \( y_{i} \)
are the energies and positions of the six selected clusters,
 had to be less than \( 10 \) cm;
\item No additional clusters with an energy \( >1.5 \) GeV were allowed within \( \pm 3 \)
ns from the event time to minimize possible accidental effects.
\end{itemize}
The longitudinal vertex positions were reconstructed analogous to (\ref{z_{t}rig}):
\begin{equation}
\label{z}
z_{vertex}=z_{LKr}-\frac{\sqrt{\sum ^{6}_{i=1}\sum ^{6}_{j>i}E_{i}E_{j}\left[ \left( x_{i}-x_{j}\right) ^{2}+\left( y_{i}-y_{j}\right) ^{2}\right] }}{m_{K}}
\end{equation}
 where \( E_{i} \) and \( x_{i} \), \( y_{i} \) are the energies and positions
of the six selected clusters.

For each combination the invariant masses \( m_{1} \), \( m_{2} \) and \( m_{3} \)
of all \( 15 \) possible photon pairings were computed using \( z_{vertex} \).
By applying a \( \chi ^{2} \) criteria, the combination and pairing most compatible
with the hypothesis that \( m_{1} \), \( m_{2} \) and \( m_{3} \) are equal
to the nominal \( \pi ^{0} \) mass were picked.

To ensure the purity of the sample, events which had one or more reconstructed
tracks in the spectrometer or more than one hit per plane in the third and fourth
drift chambers were rejected.

To improve the resolution on the Dalitz plot variables \( R^{2} \) and \( \tan (\theta ) \)
and to reduce differences in the energy resolutions and non-linearities between
data and MC which may bias the measurement of the quadratic slope parameter
\( h \), all the events were passed through a kinematical fitting procedure
with constraints. Assuming the neutral kaon coming from the \( K_{L} \) target
and a \( K\rightarrow 3\pi ^{0}\rightarrow 6\gamma  \) decay the cluster energies
and positions are adjusted to minimize \begin{equation}
\label{chi2}
\chi _{fit}^{2}=\sum _{i=1}^{6}\frac{\left( E_{i}-E_{i}^{fit}\right) ^{2}}{\sigma \left( E_{i}\right) ^{2}}+\sum _{i=1}^{6}\frac{\left( x_{i}-x_{i}^{fit}\right) ^{2}}{\sigma \left( x_{i}\right) ^{2}}+\sum _{i=1}^{6}\frac{\left( y_{i}-y_{i}^{fit}\right) ^{2}}{\sigma \left( y_{i}\right) ^{2}}
\end{equation}
 where \( \sigma \left( E_{i}\right)  \), \( \sigma \left( x_{i}\right)  \),
\( \sigma \left( y_{i}\right)  \) are given by (\ref{e_{r}esol}),(\ref{x_{r}esol})
and \( E_{i}^{fit} \), \( x_{i}^{fit} \), \( y_{i}^{fit} \) are the adjusted
energies and positions of the photons as functions of the \( 15 \) decay parameters
(the kaon energy, the vertex position, the three Euler angles of the decay plane
in the kaon rest frame, the Dalitz plot variables \( R^{2} \) and \( \theta  \)
and the 2 angles of the \( \gamma  \) directions for each of the decaying \( \pi ^{0} \)s).
While the effect of the kinematical fitting on the \( E_{K} \) and \( z_{vertex} \)
resolutions was small, the \( R^{2} \) and \( \theta  \) resolutions were
improved significantly. In particular the RMS widths of \( (R^{2}-R_{true}^{2}) \)
and \( (\theta -\theta _{true}) \) from \( 8.8\times 10^{-2} \) and \( 48 \)
mrad became \( 3.0\times 10^{-2} \) and \( 31 \) mrad, respectively.

Finally a cut \( \chi _{fit}^{2}<8.5 \) was applied to exclude the tails in
the \( \chi _{fit}^{2} \) distribution which differ for data and MC.

The numbers of data and MC events which passed the selection criteria are presented
in Table \ref{result}.

\begin{table}

\caption{\label{result}The summary of the number of accepted events and the measured
values of the Dalitz plot slope parameter \protect\( h\protect \) for the NUT
\protect\( 3\pi ^{0}\protect \), NUT \protect\( 2\pi ^{0}\protect \) and Nhodo
triggers. The errors are statistical only.\protect \\
}
{\centering \begin{tabular}{|c|c|c|c|}
\hline 
&
 NUT \( 3\pi ^{0} \)&
 NUT \( 2\pi ^{0} \)&
 Nhodo\\
\hline 
Data events&
 \( 12.43\times 10^{6} \)&
 \( 1.48\times 10^{6} \)&
 \( 0.82\times 10^{6} \)\\
\hline 
MC events&
 \( 4.76\times 10^{6} \)&
 \( 4.53\times 10^{6} \)&
 \( 2.49\times 10^{6} \)\\
\hline 
\( h\times 10^{3} \)&
 \( -6.36\pm 1.06 \)&
 \( -4.49\pm 1.88 \)&
 \( -7.53\pm 2.50 \) \\
\hline 
\end{tabular}\par}\vspace{0.5cm}\end{table}

\begin{figure}[ht]
{\par\centering \resizebox*{14.0cm}{14.0cm}{\includegraphics{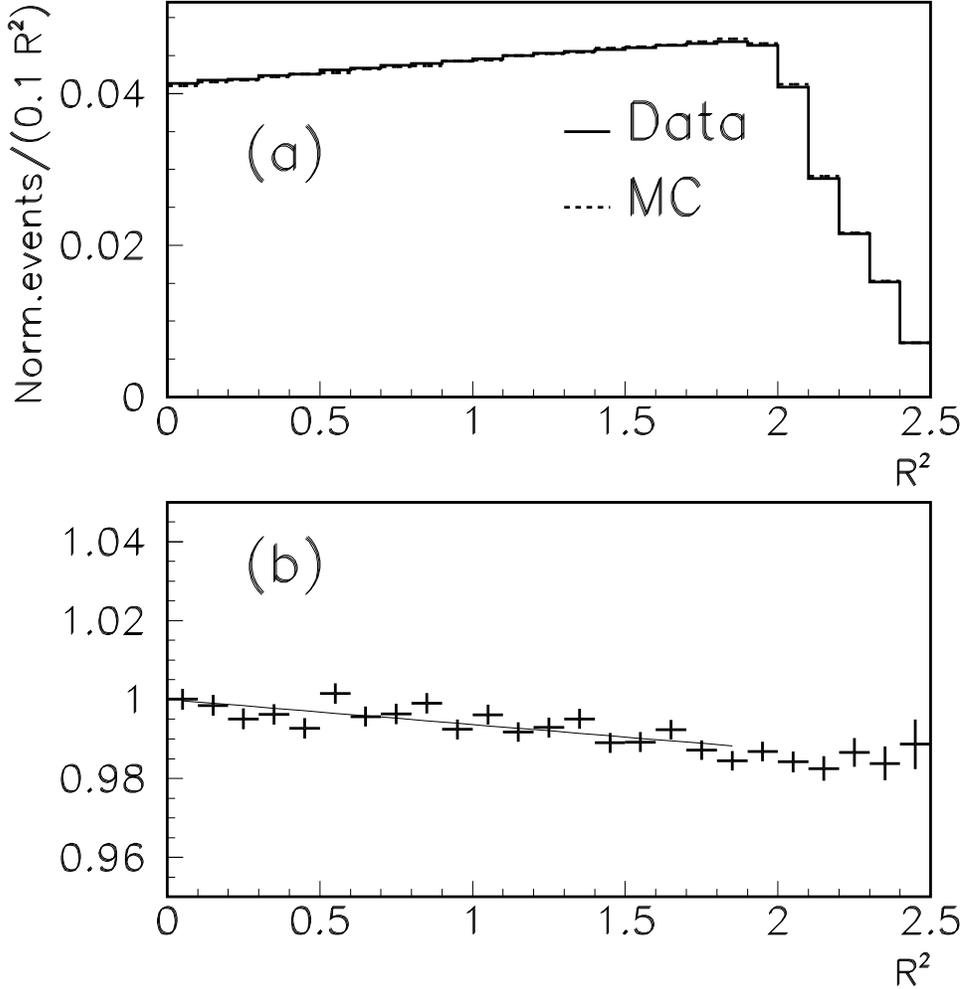}} \par}

\caption{\label{fig_r2}(a) The data (solid line) and MC (dashed line) \protect\( R^{2}\protect \)
distributions for the NUT \protect\( 3\pi ^{0}\protect \) trigger. (b) A linear
fit to the normalized at \protect\( R^{2}=0\protect \) ratio of these distributions.}
\end{figure}

To estimate the quadratic slope parameter \( h \), the \( R^{2} \) distribution
for the data events was corrected for the detector acceptance. For this purpose
about \( 8\times 10^{8} \) Monte Carlo events were generated with \( h=0 \).
Then the quadratic slope parameter \( h \) was evaluated by a linear fit to
the ratio of the data and MC \( R^{2} \) distributions (Fig. \ref{fig_r2}).
 The fit was done separately for each of the triggers using the corresponding
MC samples within appropriate decay regions. 

\begin{figure}[ht]
{\par\centering \resizebox*{14.0cm}{14.0cm}{\includegraphics{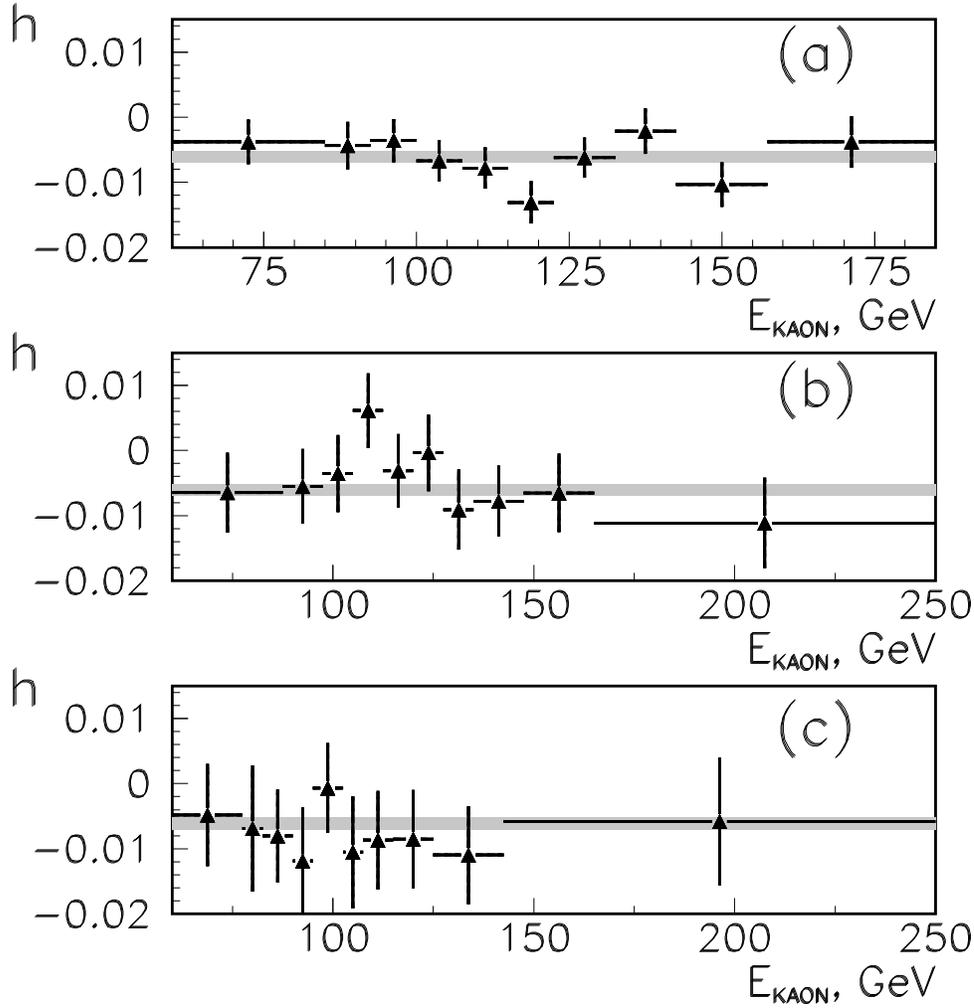}} \par}

\caption{\label{fig_h_ediv}The \protect\( K_{L}\rightarrow 3\pi ^{0}\protect \) decay
Dalitz plot slope parameter \protect\( h\protect \) in kaon energy bins for
the NUT \protect\( 3\pi ^{0}\protect \) (a), NUT \protect\( 2\pi ^{0}\protect \) (b)
and Nhodo (c) triggers, respectively. The overall result and the statistical error are shown by the shaded areas.}
\end{figure}

An important feature of the analysis was that the ratio of the data and MC \( R^{2} \)
distributions was fitted up to \( 1.9 \) (Fig. \ref{fig_r2}). This allowed
us to significantly improve the stability of the result by excluding the part
of the Dalitz plot most affected by the energy resolution and non-linearities.
Since the shape of the detector acceptance on \( R^{2} \) depends mainly on
the kaon energy, the whole analysis was done in kaon energy bins. Combining
the measured values of the Dalitz plot slope parameter \( h \) presented in
Figure \ref{fig_h_ediv}
 and Table \ref{result} we get a overall result of \( (-6.1\pm 0.9)\times 10^{-3} \).

\section*{Systematics}

Contributions of the detector acceptance estimation, the \( \chi _{fit}^{2} \)
cut, the trigger conditions and the differences in the energy resolution, non-linearities and energy
scale between the data and MC to the systematic error were considered.

In order to study possible uncertainties in the acceptance we divided the data
and MC samples upon the decay length \( \tau /\tau _{s} \), the longitudinal
vertex position \( z_{vertex} \) and the centre-of-gravity COG.
For each considered trigger sample the maximum deviation between the average value of \( h \) from Table \ref{result} and the values 
obtained averaging over the dependence of a second variable
(\( \tau /\tau _{s} \), \( z_{vertex} \), COG) was taken as systematic error
(Table \ref{acc_syst_table}). 
\begin{table}

\caption{\label{acc_syst_table}The Dalitz plot slope parameter \protect\( h\protect \)
in units of \protect\( 10^{-3}\protect \) for the different subdivisions of
the data and MC.\protect \\
}
{\centering \begin{tabular}{|c|c|c|c|}
\hline 
&
 NUT \( 3\pi ^{0} \)&
 NUT \( 2\pi ^{0} \)&
 Nhodo\\
\hline 
In \( z_{vertex} \)bins &
-6.50 &
-4.48 &
-6.72\\
\( \chi ^{2}/n.d.f \)&
14.3/9 &
7.4/10 &
6.9/10\\
\hline 
In \( \tau /\tau _{s} \)bins &
-6.43 &
-4.58 &
-7.70\\
\( \chi ^{2}/n.d.f \)&
14.6/9 &
8.9/9 &
7.2/9\\
\hline 
In \( COG \)bins &
-6.38 &
-4.64 &
-7.56\\
\( \chi ^{2}/n.d.f \)&
12.4/9 &
11.6/9 &
5.7/9\\
\hline 
\( \Delta h \)&
0.14 &
0.16 &
1.0 \\
\hline 
\end{tabular}\par}\vspace{0.5cm}\end{table}

\begin{figure}[ht]
{\par\centering \resizebox*{14.0cm}{14.0cm}{\includegraphics{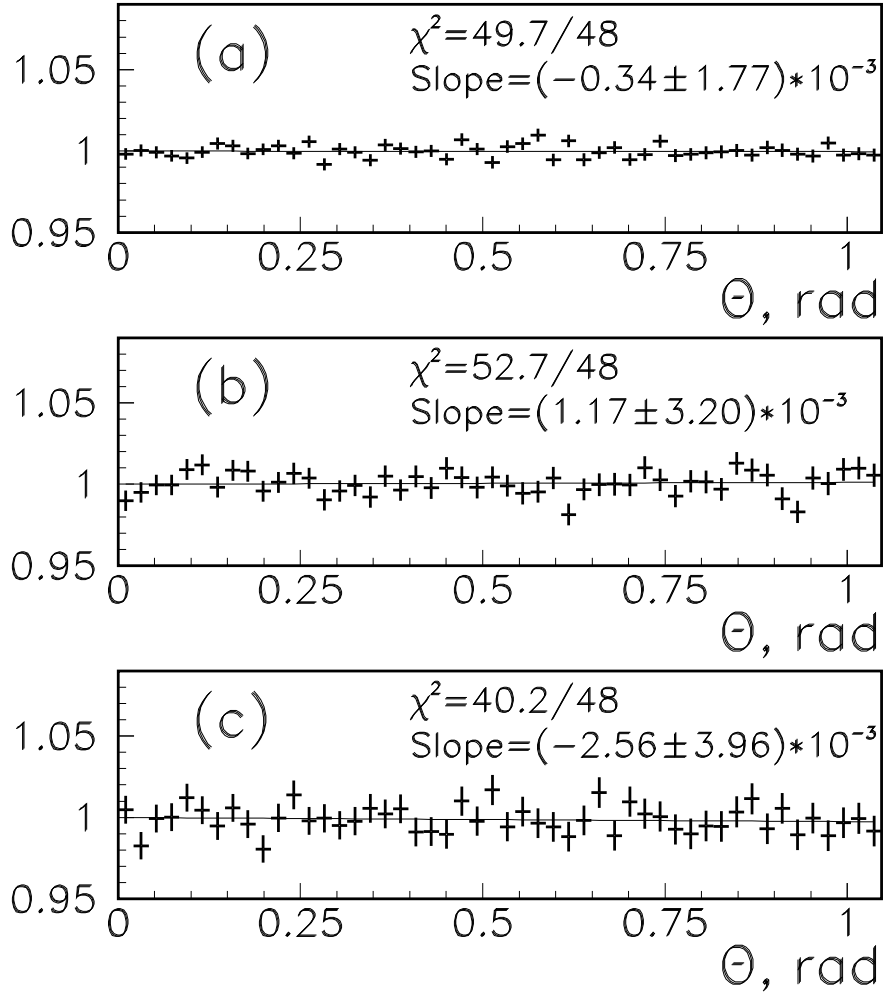}} \par}

\caption{\label{fig_theta}The normalized ratios of the data and MC \protect\( \theta \protect \)
distributions for the NUT \protect\( 3\pi ^{0}\protect \) (a), NUT \protect\( 2\pi ^{0}\protect \)
(b) and Nhodo (c) triggers. The ratios were fitted with \protect\( f(\theta )=(1+Slope\times \theta )\protect \).}
\end{figure}

An important check of the reliability of the MC was done by comparing the data
and MC Dalitz plot variable \( \theta  \) (Fig. \ref{fig_theta}).
 A linear fit to the ratio of the data and MC \( \theta  \) distributions clearly
points to a very good agreement between experimental data and the simulation.

The second considered source of systematics was the difference between the data
and MC cluster energy resolution. Using the measured and kinematically adjusted
cluster energies we established a strong control over the resolutions seen in
the data and the MC. The differences in the resolutions were adjusted by an
additional smearing to the cluster energies in the MC. The maximum change in
\( h \) after these adjustments was found to be \( \pm 0.20\times 10^{-3} \)
and was taken as a systematic error. A similar approach was applied in the
study of the systematics due to the energy non-linearities leading to a \( \pm 0.17\times 10^{-3} \)
systematic error on the \( h \). To take into account small non-gaussian tails
in the cluster energy resolution seen in the data, a simplified parameterization
was used to add the tails to the MC events. The effect of the tails was found
to be negligible. Changing the energy scale in the MC by \( \pm 1\times 10^{-3} \)
we got a shift in \( h \) of \( \pm 0.17\times 10^{-3} \).

The influence of the trigger conditions was studied by subdividing the data and 
MC in \( \tau /\tau _{s} \) bins. Although that for the NUT \( 2\pi ^{0} \) 
and Nhodo trigger samples no systematic effect was found, the result for
NUT \( 3\pi ^{0} \) trigger sample was affected by a small NUT inefficient 
region near the \( \tau ^{trig} /\tau _{s} \) cut. The data and MC were 
processed excluding this region and the change in
the result of \( 0.1\times 10^{-3} \) gave us a systematic error due to the
trigger.
%Since the measured \( h \) values for the three used triggers are in agreement,
%the systematic error due to the trigger is supposed to be small. Nevertheless,
%looking at the NUT \( 3\pi ^{0} \) data and MC divided in \( \tau /\tau _{s} \)
%bins we found a rather small effect at \( \tau /\tau _{s} \) from \( 3 \)
%to \( 3.5 \). The effect was identified as NUT inefficiency near the \( \tau /\tau _{s} \)
%cut. The data and MC were processed excluding this region and the change in
%the result of \( 0.1\times 10^{-3} \) gave us a systematic error due to the
%trigger. For the two other trigger samples no effect was seen up to \( 6\times \tau /\tau _{s} \).

The stability of the result upon the \( \chi _{fit}^{2} \) cut was checked
by loosening it to \( 17 \). No statistically significant effect on \( h \)
was established. Furthermore, the result was found to be stable upon different
cuts on the clusters energies and positions in LKr.

No influence of a possible background contamination from \( \gamma  \) conversions
and Dalitz decays was observed.

All the sources of systematic error were added quadratically and yield a total
systematic error of \( 0.5\times 10^{-3} \).

In addition, a second independent analysis has been performed yielding a consistent
result.

\section*{Conclusions}

Our result for the Dalitz plot slope parameter in the \( K_{L}\rightarrow 3\pi ^{0} \)
decay is \( h=(-6.1\pm 0.9_{stat}\pm 0.5_{syst})\times 10^{-3} \), where the
errors are statistical and systematic, respectively. This result is in an
agreement with the value of \( (-12\pm 4)\times 10^{-3} \) predicted in \cite{Kambor1}.
The previous measurement of \( \left( -3.3\pm 1.1_{stat}\pm 0.7_{syst}\right) \times 10^{-3} \)
performed by the E731 experiment \cite{E731}, is comparable with the result
presented in this paper.

\section*{Acknowledgements}

We would like to thank the technical staff of the participating laboratories,
universities and affiliated computing centres for their support and co-operation.

\end{document}